\documentclass[twocolumn,showpacs,prl,aps,final,letterpaper,nobibnotes]{revtex4}
\usepackage{bm,amsmath,amssymb}
\usepackage[dvips]{graphicx}

\begin{document}

\title{Gravitational Lensing and Gravitomagnetic Time Delay}

\author{Ignazio Ciufolini}
\affiliation{Dip.
Ingegneria dell'Innovazione Universit\`a di Lecce \\ Via
Monteroni, 73100 Lecce, Italy}
\author{Franco Ricci}
\affiliation{Dip. Fisica Universit\`a di
Roma ``La Sapienza'' \\ Piaz.le Aldo Moro 5, 00185 Roma, Italy}

\date{\today}

\begin{abstract}
We derive the delay in travel time of photons due to the spin of
a body both inside a rotating shell and outside a rotating body.
We then show that this time delay by the spin of an astrophysical
object might be detected in different images of the same source
by gravitational lensing; it might be relevant in the
determination of the Hubble constant using accurate measurements
of the time delay between the images of some gravitational lens
systems. The measurement of the spin-time-delay might also provide
a further observable to estimate the dark matter content in
galaxies, clusters, or super-clusters of galaxies.

\end{abstract}


\maketitle

One of the basic confirmations of Einstein general relativity is
the measurement of the deflection of path of electromagnetic
waves propagating near a mass due to the spacetime curvature
generated by a central body \cite{wil1,ciuw}. To-day this effect
is reported to agree with its general relativistic value with
accuracy of about $1.4$ parts in $10^4$, by using VLBI, Very Long
Baseline radio Interferometry to observe the gravitational
deflection of radio waves from quasars and radio galaxies
produced by the Sun \cite{VLBI}. The VLBI measurement constrains
the parameter $\omega$ of scalar-tensor theories to be larger
than about 3500 \cite{wil2}. Depending from the relative
position, distance and alignment of source, observer and
deflecting mass, several images of the same source may be
observed. This phenomenon is the well known gravitational
lensing. Several examples of gravitational lensing have been
discovered; a well known case is the Gravitational Lens G2237 +
0305, or Einstein Cross, where the light bending produces four
images of the same quasar as observed from Earth \cite{GL,GL'}.

The other well known general relativistic effect due to the
spacetime curvature generated by a central body is the delay in
the travel time of electromagnetic waves propagating near the
central mass, or Sha\-piro time delay \cite{sha,sha',sha''}; this
effect is to-day tested with accuracy of about $10^{-3}$ by
observing the delay of  radio waves propagating near the Sun with
active reflection using transponders on the Viking spacecraft
orbiting Mars or on its surface.

All these phenomena are due to a static mass and can be derived
using the static Schwarzschild metric. If an axially symmetric
body has a steady rotation around its symmetry axis an external
solution is the stationary Kerr metric. In the weak-field and
slow-motion limit, the Kerr metric is, in Boyer-Lindquist
coordinates, the Lense-Thirring metric \cite{lent, ciuw}.
Einstein's general theory of relativity \cite{ciuw} predicts that
when a clock  co-rotates arbitrarily slow around a spinning body
and returns to its starting point, it finds itself advanced
relative to a clock kept there at "rest" (in respect to "distant
stars"), and a counter-rotating clock finds itself retarded
relative to the clock at rest \cite{ciuw,ciur}. Indeed,
synchronization of clocks all around a closed path near a
spinning body is not possible, and light co-rotating around a
spinning body would take less time to return to a fixed point
than light rotating in the opposite direction. Similarly, the
orbital period of a particle co-rotating around a spinning body
would be longer than the orbital period of a particle
counter-rotating on the same orbit. Furthermore, an orbiting
particle around a spinning body will have its orbital plane
"dragged" around the spinning body in the same sense as the
rotation of the body, and small gyroscopes that determine the
axes of a local, freely falling, inertial frame, where "locally"
the gravitational field is "unobservable," will rotate in respect
to "distant stars" because of the rotation of the body. This
phenomenon--called "dragging of inertial frames" or, "frame
dragging," as Einstein named it--is also known as Lense-Thirring
effect. The Lense-Thirring effect has been observed in the orbits
of the  LAGEOS satellites \cite{ciup}. In general relativity, all
these phenomena are the result of the rotation of the central
mass.

However, Einstein's gravitational theory predicts peculiar
phenomena also inside a rotating shell. In a well known paper of
1918, Thirring published a solution of Einstein's field equation
representing the metric inside a rotating shell to first order in
$\frac{M}{R}$, mass over radius of the shell, and to first order
in $\omega$, angular velocity of the shell \cite{thi}. In 1966
Brill and Cohen derived the metric inside a shell with arbitrary
mass, this solution is a lowest order series expansion in the
angular velocity of the spherical shell on the Scharzschild
background of any mass, valid both inside and outside the shell
\cite{bric}. An extension of the Brill-Cohen results to higher
orders on $\omega$ was then published in 1985 by Pfister and
Braun \cite{pfib}. However, the exact solution representing the
spacetime geometry inside a shell with arbitrary mass and
rotating with arbitrary angular velocity is still unknown. In the
following we consider only the weak-field and slow-motion metric
derived by Thirring.
\begin{figure}
\fbox{\includegraphics[width=\linewidth]{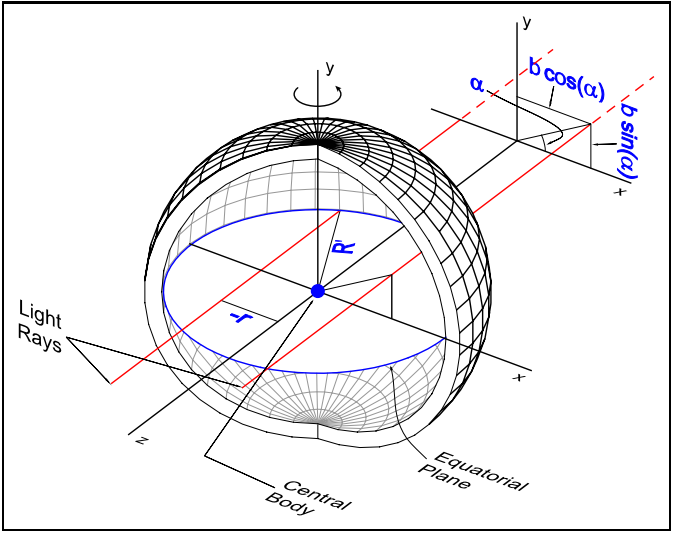}}
\caption{Geometry of light rays propagating inside a rotating
shell of radius $R'$ and around a central body.} \label{fig1}
\end{figure}

The gravitomagnetic potential, i.e. the non-diagonal part of the
metric in standard isotropic PPN coordinates \cite{wil1},
$h_{0i}$, at the post-Newtonian order, outside a stationary
rotating body with angular momentum $\mathbf{J}$, is \cite{ciuw}:
\begin{equation} \label{1}
{\mathbf{h^{ext}}}  ({\mathbf{X}}) ~ \cong \frac{- 2 ~
{{\mathbf{J}}  \times  {\mathbf{X}}}}{{\vert {\mathbf{X}} \vert
^3}},
\end{equation}
where $\mathbf{X}$ is the position vector. If
$\mathbf{J}=(0,0,J)$, in spherical coordinates: $h^{ext}_{0
\phi}\cong  -  {\frac{2 J}{r}} sin^2 \theta$. Inside a thin shell
of total mass $M$ and radius $R$, with stationary rotation around
a Z-axis with small angular velocity $\bm{\omega}$, we have
\cite{ciuw}:
\begin{equation}\label{2}
{\mathbf{h^{int}}} ({\mathbf{X}}) = - \frac{4M}{3R}\,
{\bm{\omega}} \times {\mathbf{X}} = \left({\frac{4M}{3R} }
\,\omega Y,  - {\frac{4M}{3R} }  \,\omega X,
 0 \right)
\end{equation}

Let us derive time delay and deflection of electromagnetic waves
due to spin and quadrupole moment of the central body; we keep
only post-Newtonian terms in the following derivation. Deflection
by spin was analyzed in \cite{iba}, and time delay by spin is
considered in \cite{kli,kop}. Strong gravitational lensing due to
a Schwarzschild black hole is treated in \cite{ellis}. In the
following, the quasi-Cartesian coordinate system $(x,y,z)$ is a
system of isotropic coordinates with the z-axis directed towards
the observer on Earth, while $(X,Y,Z)$ is the standard PPN system
of isotropic coordinates attached to the deflecting body. To
relate the coordinates $(x,y,z)$ and $(X,Y,Z)$ we use the standard
Euler's angles $(\phi,\beta,\gamma)$. If the deflecting body is
symmetric with respect to $Z$, the shape of the body is invariant
for rotations of $\phi$. We can thus choose $\phi = 0$. Since we
analyze the behaviour of photons, moving with speed $c \equiv 1$,
we neglect the $U^2$ terms in the post Newtonian metric, where
$U$ is the Newtonian potential. In the new coordinate system
$(x,y,z)$ we then have:
\begin{eqnarray}
ds^2  &=& \left( { - 1 + 2 U} \right) dt^2 +
\left( {1 + 2 U} \right)\delta _{ij} dx^i\,dx^j\nonumber\\
&&+\frac{{4\,J}} {{r^3 }}\left( {y\cos \beta  - z\cos \gamma \sin
\beta } \right)dt\,dx \nonumber\\
&&- \frac{{4\,J}}{{r^3 }} \left( {x\cos \beta  - z\sin \gamma
\sin \beta } \right)dt\,dy \nonumber\\
&&+\frac{{4\,J}}{{r^3 }}\left( {x\cos \gamma \sin \beta  - y\sin
\gamma \sin \beta } \right)dt\,dz \label{3}
\end{eqnarray}
in $U$, for simplicity, we have only included mass and quadrupole
moment of the deflecting body.

We have chosen the quasi-Cartesian coordinates so that the
emitting and the deflecting bodies have the same $x$ and $y$
coordinates, (see fig.\ref{fig1}), but a different $z$
coordinate: source, lens and observer are aligned. We have chosen
this simple configuration since, in this paper, we are only
interested to analyze the time delay due to spin. However, there
is an additional time delay, called geometrical time delay
\cite{sch}, due to the different geometrical path followed by
different rays. Depending on the geometry of the system, this
additional term may be very large and may be the main source of
time delay. However, if we compare the time delay of photons that
follow the same geometrical path we can neglect the geometrical
time delay, as in the case of two light rays with the same impact
parameter but on different sides of the deflecting object. For of
a small deflection angle, the contribution to the travel time
delay from the different path length traveled, due to the small
deflection, is of the order of $U^2$ \cite{ciuw} and, depending
on the geometrical configuration considered, this delay may need
to be included in the total time delay. In a following paper we
shall analyze higher order time delay and compare it with the
gravitomagnetic time delay. Here, for simplicity, we neglect any
geometrical time delay. Thus, from the condition of null arc
length, $ds^2 \, = 0$, solving with respect to $dt$, we have at
the lowest order in $U$ and $g_{0z}$: $dt \simeq { g_{0z} }dz \mp
\left({1 + 2 U} \right)dz$. Thus, by integrating $dt$ from $-z_1$
to $z_2$ corresponding, respectively, to the position of source
and observer, if $z_1 \simeq z_2 \equiv \overline{z} \gg b$ (see
fig. \ref{fig1}) we get:
\begin{eqnarray}
\Delta T &=& 2\,\bar z + 4\,M\ln \left( {2\frac{{\bar z}} {b}}
\right) +  \frac{{4\,J\cos \,(\alpha + \gamma )\,\sin \beta }}
{b}\nonumber\\
&&\frac{{2\,M\,R^2 \,J_2 \cos 2\left( {\alpha  + \gamma }
\right)\,\sin \beta ^2 }} {{b^2 }}  \label{4}
\end{eqnarray}
In this expression the first term is the time that a radio pulse
takes to travel from the source to Earth in the absence of a
central mass: $M=0$; the second term is the Shapiro time delay;
the third one is the gravitomagnetic time delay and the last term
is the additional time delay due to the quadrupole moment, $J_2$,
of the deflecting body.

Together with the positive spin-time-delay of a counter-rotating
photon relative to a co-rotating photon, there is a negative
deflection of the path of the counter-rotating photon, and a
positive deflection of the co-rotating photon, due to the spin of
the lens. This negative deflection gives rise to a negative time
delay, due to the decrease in the path travelled, and to a
positive time delay due to the increase in the standard Shapiro
delay by the mass of the lens due to the decrease of the distance
from the central lens. These two additional contributions,
negative geometrical delay and positive Shapiro delay, are equal
and opposite and cancel out so that the only remaining effect is
the positive spin-time-delay of the counter-rotating photon and
the negative spin-time-delay  of  the co-rotating photon.

To write the deflection of electromagnetic waves due to spin and
quadrupole moment of the deflecting body we use the geodesic
equation in the weak field approximation. When source, deflecting
body and observer are aligned, after some calculations we then
get \cite{ciur}: $\delta =  - \frac{4M} {b} - \frac{{4J\sin
\beta}} {{b^2 }} -  \frac{{4J_2 M\,R^2 \sin ^2 \beta}} {{b^3 }}$;
where the first term is the standard deflection by a spherical
object of mass $M$, the second term is the deflection by the
angular momentum, $J$, and the third one the additional
deflection due to the quadrupole moment, $J_2$, of the central
body.

Let us now study the possibility of measuring the spin-time-delay
and determining the angular momentum $J$ of the central
deflecting body by measuring total time delay. We consider a
simple case in which the source is behind the lens and there are
three light rays with the same impact parameter $b$, propagating
along one axis. We assume to be able to measure, or determine,
the following quantities: total time delay between the three
rays, $\Delta T_{12}$ and $\Delta T_{13}$; deflection angles
$\delta _{1}$, $\delta _{2}$ and $\delta _{3}$; equatorial radius
$R$ of the deflecting body and distances of source and lens from
the observer. In this way we are able to determine the angle
$\alpha$ for each light beam and the impact parameter $b$ (see
fig. \ref{fig1}), and we can write a system in which the only
unknown quantities are: angular momentum, $J$, quadrupole moment,
$J_2$, mass, $M$, and Euler's angle $\beta$ and $\gamma$. Solving
this system we can, in principle, determine the time delay due to
the angular moment $J$ and the other unknown quantities. Detailed
calculations are given in \cite{ciur}. Of course, for other
configurations in which the source is not exactly aligned with
lens and observer, the difference in path traveled and the
corresponding difference in Shapiro time delay can be the main
source of relative time delay; one would then need to model and
remove these delays between the different images on the basis of
the observed geometry of the system. Nevertheless, in special
cases, for example if we observe four images of the source and if
the angle $\alpha$ of each deflected ray differs by about $\pi$,
such as in the Einstein Cross, we can directly eliminate the time
delay due to the quadrupole moment and thus determine the
spin-time-delay.

Let us now calculate the time delay due to the spin of some
astrophysical sources. For the sun ($M_\odot = 1.477$ km,
$R_\odot = 6.96\cdot 10^5$ km and ${J_2}_\odot \cong 1.7\cdot
10^{-7}$ \cite{ciuw}), by considering two light rays with impact
parameters $b \cong R_\odot$ and $- b$, and, for simplicity, with
$\gamma=0$ and $\beta=\frac{\pi}{2}$, the gravitomagnetic and
quadrupole-moment time delays, according to (\ref{4}), are:
$\Delta T_{12}^J=\frac{8J}{b}=1.54\cdot 10^{-11}\,\text{sec}$, and
$\Delta T_{12}^{J_{2}}=4\frac{J_{2}MR^{2}}{b^2}=3.35\cdot
10^{-12}\,\text{sec}$. The time delay due to the Sun spin could
then, in principle, be measured using an interferometer at a
distance of about $8\cdot 10^{10}$ km, by detecting by
gravitational lensing photons emitted by a laser on the side of
the Sun opposite the detector, and travelling on opposite sides
of the Sun. To derive the time delay due to the lensing galaxy of
the Einstein cross \cite{GL,GL'} we assume a simple model for
rotation and shape of the central object. Details about this
model can be found in \cite{cha}. The angular separation between
the four images is about $0.9''$, corresponding to a radius of
closest approach of about $R\simeq 650 h^{-1}_{75}$ pc, and the
mass inside a shell with this radius $R$ is $\sim 1.4\cdot
10^{10}\,h^{-1}_{75}\,M_{\odot}$ \cite{GL'}. Let us assume:
$J_2\simeq 0.1$ and $J\simeq 10^{23}\, km^2\, h^{-2}_{75}$, we
then have $\Delta T_{12}^J=\frac{8J}{b}\simeq 4\,min$, and $\Delta
T_{12}^{J_{2}}=4\frac{J_{2}M\,R^{2}}{b^2}\simeq8\,hr$. Thus, at
least in principle, one could measure the time delay due to the
spin of the lensing galaxy by removing the larger
quadrupole-moment time delay by the previously described method;
of course, as in the case of the Sun, one should be able to
accurately enough model and remove all the other delays, due to
other physical effects, from the observed time delays between the
images. As a third example we consider the relative time delay of
photons due to the spin of a typical cluster of galaxies; the
precise calculations are shown in a following paper. We consider
a cluster of galaxies of mass $M_C \cong 10^{14} M_{\odot}\,$,
radius $R_C \cong 5 \,$ Mpc and angular velocity $\omega_C \cong
10^{-18} \text{s}^{-1}$; depending on the geometry of the system
and on the path followed by the photons, we then find relative
time delays ranging from a few minutes to several days
\cite{ciur}.

Let us now analyze the time delay in the travel time of photons
propagating inside a rotating shell with
$\bm{\omega}=(0,0,\omega)$. Inside the shell it is not possible to
synchronize clocks all around a closed path. Indeed, if we
consider a clock co-rotating very slowly  along a circular path
with radius $r$, when back to its starting point it is advanced
with respect to a clock kept there at rest (in respect to distant
stars). The difference between the time read by the co-rotating
clock and the clock at rest is equal to:
\begin{equation}
\delta T = - \oint {\frac{{g_{0i} }} {{\sqrt{-g_{00}}
}}dx^i}=\frac{{8M}} {3R}\pi \omega r^2 \label{5a}
\end{equation}
For a shell with finite thickness we just integrate (\ref{5a})
from the smaller radius to the larger one.

Let us now consider a co-rotating photon traveling with an impact
parameter $r$ on the equatorial plane of a galaxy (see fig.
\ref{fig1}). The time delay due to the rotation of the external
mass for every infinitesimal shell with mass $dm=4\pi \rho R'^2
dR'$ and radius $R'\geq \left| r \right|$, is \cite{ciur}:
\begin{equation}
\Delta t_{dm} = \int_{ - \sqrt {R'^2 - r^2 } }^{\sqrt {R'^2  -
r^2 } } {\left(h_{0x}\right) dx} = \frac{{8\,dm}} {3}\omega
r\frac{{\sqrt {R'^2 - r^2 } }} {{R'}}\label{5b}
\end{equation}
By integrating the second term in this expression from ${\left| r
\right|}$ to the external shell radius $R$, we have:
\begin{equation}\label{6}
\Delta T =\frac{{32\pi }} {3}\omega r\int_{\left| r \right|}^R
{\rho R'} \sqrt {R'^2  - r^2 } dR'
\end{equation}
This is the time delay due to the spin of the whole rotating mass
of the external shell. From this formula we can easily calculate
the relative time delay between two photons traveling on the
equatorial plane of a rotating shell, with impact parameters
$r_1$ and $r_2$.

Finally, let us calculate the time delay corresponding to some
astrophysical configurations. In the case of the "Einstein cross"
\cite{GL}, we assume, to get an order of magnitude of the effect,
that the lensing galaxy has an external radius $R\simeq 5\,kpc$;
after some calculations based on the model given in ref.
\cite{cha}, the relative time delay of two photons traveling at a
distance of $r_1\simeq650\,pc$ and $r_2\simeq - 650\,pc$ from the
center, using (\ref{6}) in the case $r _1\simeq - r_2$, is:
$\Delta T\simeq 30\,\text{min}$. If the lensing galaxy is inside
a rotating cluster, or super-cluster, to get an order of
magnitude of the time delay, due to the spin of the mass rotating
around the deflecting galaxy, we use typical super-cluster
parameters: total mass $M=10^{15}\,M_\odot$, radius $R=70\,
\text{Mpc}$ angular velocity $\omega=2 \cdot 10^{-18}\, s^{-1}$
\cite{abe}. If the galaxy is in the center of the cluster and
light rays have impact parameters $r_1 \simeq 15\, kpc$ and $r_2
\simeq - 15\, kpc$ (of the order of the Milky Way radius), the
time delay, applying  formula (\ref{6}) in the case $r_1 \simeq -
r_2$ and $\rho= costant$, is: $\Delta t\simeq 1 \, day$.

If the lensing galaxy is not in the center of the
cluster but at a distance $r = a R$ from the center, with $0 \leq
a \leq 1$ and $R$ radius of the cluster, by integrating (\ref{6})
between $r = a R$ and $R$, when $r_1 \simeq r_2$ we have: $\Delta
t = \frac{32\pi}{9} \omega ({r_1  - r_2}) \rho {(1 - a^2)^{1/2}}
(1 - 4 a^2) R^3$. Thus, if the lensing galaxy is at a distance of
10 Mpc from the center of the cluster, the relative time delay
due to the spin of the external rotating mass between two photons
with $(r_1-r_2)\simeq 30\, kpc$, is: $\Delta T\simeq 0.9 \, day$.

Promising candidates to observe time delay due to spin are
systems of the type of the gravitational lens B0218+357
\cite{big}, where the separation angle between the images is so
small that also the delay between the images is very small. In
B0218+357 the separation angle between the two images is 335
milliarcsec and the time delay is about 10.5 days. In such
systems, the time delay due to the spin of the external mass, or
of the central object, might be comparable in size to the total
delay. In addition, in the system B0218+357 is observed an
Einstein ring the diameter of which is the same as the separation
of the images. In such configurations, the Einstein ring can
provide strong constraints on the mass distribution in the lens;
this, in turn, can be used in order to separate the time delays
due to a mass distribution non-symmetrical with respect to the
images. The accurate measurement of the delay between the images
of some gravitational lens systems is used as a method to provide
estimates of the Hubble constant, the time delay due to spin
might then be relevant in the corresponding modeling of the delay
in the images. The measurement of time delay is possible, in the
case of B0218+357, because the source is a strongly variable
radio object, thus one can determine the delay in the variations
of the images. In this system it is possible to observe clear
variations in total flux density, percentage polarization and
polarization position angle at two frequencies. For B0218+357 the
measured delay is 10.5 $\pm$ 0.4 day \cite{big}. Therefore. since
the present measurement uncertainty in the lensing time delay is
of the order of 0.5 day \cite{big2}, the gravitomagnetic time
delay might already be observable.

In conclusion, we have derived and studied the "{\it
spin-time-delay}" in the travel time of photons propagating near
a rotating body, or inside a rotating shell due to the angular
momentum We found that there may be an appreciable time delay due
to the spin of the body, or shell, thus spin-time-delay must be
taken into account in the modeling of relative time delays of the
images of a source observed at a far point by gravitational
lensing. This effect is due to the propagation of the photons in
opposite directions with respect to the direction of the spin of
the body, or shell. If other time delays can be accurately enough
modeled and removed from the observations, the larger relative
delay due to the quadrupole moment of the lensing body can be
removed, for some configurations of the images, by using special
combinations of the observables; thus, one could directly measure
the spin-time-delay due to the gravitomagnetic field of the
lensing body. In order to estimate the relevance of the
spin-time-delay in some real astrophysical configurations, we
have considered some possible astrophysical cases. We analyzed
the relative time delay in the gravitational lensing images
caused by a typical rotating galaxy, or cluster of galaxies. We
then analyzed the relative spin-time-delay when the path of
photons is inside a galaxy, a cluster, or super-cluster of
galaxies rotating around the deflecting body; this effect should
be large enough to be detected from Earth. The measurement of the
spin-time-delay, due to the angular momentum of the external
massive rotating shell, might be a further observable for the
determination of the total mass-energy of the external body, i.e.
of the dark matter of galaxies, clusters and super-clusters of
galaxies. Indeed, by measuring the spin-time-delay one can
determine the total angular momentum of the rotating body and
thus, by estimating the contribution of the visible part, one can
determine its dark-matter content. The estimates presented in
this paper are preliminary because we need to analyze the
spin-time-delay in the case of some particular, known,
gravitational-lensing images; furthermore, we need to estimate the
size and the possibility of modeling other sources of time delay
in these known systems. Nevertheless, depending on the geometry of
the astrophysical system considered, the relative spin-time-delay
can be a quite large effect.

\bibliography{ciufolini271202}

\end{document}